\documentstyle[11pt,paspconf,psfig]{article}
\def\ref{{\reference}}

\newdimen\thicksize
\newdimen\thinsize
\def\th{\thinspace}

\def\qquad{\quad\quad}

\def\frac#1#2{{\textstyle{ #1 \over #2}}}

\def\etal{{\sl et al.\ }}

\def\at{{\rm\char'100}}
\def\ni{\noindent}

\def\lgt{ \raise4pt \hbox{$<$}\kern-9pt\lower1.5pt \hbox{$>$}}
\def\glt{\raise4pt \hbox{$<$}\kern-9pt\lower1.5pt\hbox{$>$}}

\def\approxgt{\raise3pt\hbox{${\scriptstyle>}$}
           \kern-6pt\lower1.1pt\hbox{${\scriptstyle\sim}$}}
\def\approxlt{\raise3pt\hbox{${\scriptstyle<}$}
           \kern-6pt\lower1.1pt\hbox{${\scriptstyle\sim}$}}

\long\def\jumpover#1{{}}

\begin{document}

\title{Nonlinear Pulsations}

\author{J.~R.~Buchler{\altaffilmark{1}}}
\affil{Physics Department, University of Florida,
Gainesville, FL 32611, USA }

\altaffiltext{1}{e--mail: buchler\at phys.ufl.edu}

\begin{abstract}
 We review some of the recent advances in nonlinear pulsation theory, but also
insist on some of the major extant shortcomings.
 \end{abstract}


\keywords{Nonlinear pulsations, variable stars, Cepheids, RR~Lyrae, W~Virginis,
RV~Tauri stars}

\vspace{-10pt}
\section{Introduction}
\vspace{-2pt}

It is a great pleasure for me to dedicate this lecture to Art Cox whose
contributions to nonlinear pulsations go back to the very pioneering years of
numerical hydrodynamics some 35 years ago.  It is fair to say that there
isn't a type of star that Art hasn't attempted to model.

It is of course impossible to cover in this short review all the topics of
interest to nonlinear stellar pulsations and I will have to make a selection
that necessary reflects my biases.  There has been a huge amount of work done
on stellar pulsations and I also apologize upfront for many important
omissions.  Extensive references are provided in the excellent reviews of
Gautschy and Saio (1995, hereafter GS).

There are basically two approaches to nonlinear stellar pulsations that are
complementary in some ways.  The first, and oldest, is numerical hydrodynamics.
While this is a 'brute force' approach, it has the advantage that with
state-of-the-art physical input and numerical methods it can yield accurate
information about the nonlinear pulsations of individual stellar models.  The
second approach is the amplitude equation formalism, and it is perhaps of a
more fundamental nature (e.g. Buchler 1988).  It gives a broad overview of the
possible behavior, such as modal selection (bifurcations in modern language)
and the effects of resonances (Buchler 1993).  We note that currently this is
the {\sl only} tool with which we can understand nonlinear {\sl nonradial}
pulsations.

About 25 years ago John Cox (1975) felt that "overall, pulsation theory and
its applications are in a fairly satisfactory state, except for a few
disturbing problems".  In the intervening time, of course, a good deal of
progress has been made.  However, there remain some "disturbing problems", and
in fact some new ones have appeared recently.

Perhaps the most important progress came from outside pulsation theory, namely
from a revision of the stellar opacities (Iglesias \& Rogers 1992, Seaton,
Kwan, Mihalas \& Pradhan 1994), and it essentially solved two longstanding
Cepheid problems.  First, the so-called Cepheid bump mass problem (e.g. Art Cox
1980) essentially disappeared and the agreement with observation appears now to
be quite satisfactory for the Galactic Cepheids.  (Moskalik, Buchler \& Marom
1992, Kanbur \& Simon 1993).  Second, the beta cephei models finally became
linearly unstable (cf. GS).

The very first numerical Lagrangean hydrodynamical computations of
\hyphenation{Ceph-eid} Cepheid variables were already apologetic about the poor
resolution of the partial hydrogen ionization front during the pulsation.  In
the late 70s Castor, Davis \& Davison (1977) developed the first code that
could track the moving sharp features and Aikawa \& Simon (1983) started to use
it systematically.  More recently, taking advantage of new developments in
computational physics and of faster computers, several groups have developed
more flexible adaptive codes (Gehmeyr \& Winkler 1992, Dorfi \& Feuchtinger
1991, Buchler, Kollath \& Marom 1997).  With these codes it is now possible to
obtain a good spatial resolution that satisfactorily resolves all shocks and
ionization fronts and achieves a much enhanced numerical accuracy of the
pulsation.  The most striking improvement is in the smoothness of the
lightcurves and radial velocity curves.  Fortunately, though, we do have to
discard the Lagrangean results since quantities such as Fourier decomposition
parameters are not substantially different from those obtained with Lagrangean
codes.

However, instead of congratulating ourselves on these and other successes,
it is perhaps more useful to dwell on the "disturbing problems".

\vspace{-8pt}
\section{'Disturbing Problems'}
\vspace{-5pt}

\subsection{Low metallicity Cepheids:}
\vspace{-2pt}

The microlensing projects have provided us with a large treasure trove on
variable stars in the Magellanic Clouds, and since these galaxies have been
found to be metal deficient compared to the Galaxy the new observations have
considerably enlarged our data base, especially for the Cepheids since they
should be strongly affected by metallicity content.

The Fourier decomposition parameters of the fundamental Cepheid
\hyphenation{light-curves} lightcurves in the SMC (Beaulieu \& Sasselov 1997)
and the LMC (Beaulieu et al. 1995, Welch et al. 1995) indicate that the
$\phi_{21}$ phase progression is very similar to that of the Galaxy, although
it may be shifted by $\pm$ 1 or perhaps 2 days in period.  However, the size of
the excursion in $\phi_{21}$ in the resonance region is essentially the same as
in the Galaxy.

A comparison of the {\sl linear} bump Cepheid models with mass--luminosity
relations derived from evolutionary computations show up an irreconcilable
difference with the observations (Buchler, Koll\'ath, Beaulieu \& Goupil 1996).
Furthermore the {\sl nonlinear} calculation of low Z Cepheid model pulsations
give $\phi_{21}$ in which the size of the excursion in the 10 day resonance
region almost vanishes as one goes to Z values of 0.005 (Fig.~1) (see also
poster by Goupil).

As far as beat Cepheids are concerned, the hope had been that the observed
period ratios could give a powerful constraint on the stellar model parameters.
While globally, it might appear that the observed period ratios of the
Galactic, LMC and SMC beat Cepheids are in agreement with the linear models
obtained with the new opacities, when looked at in detail, i.e. by considering
individual stars, the agreement is no longer as good.  More seriously,
(unknown) nonlinear period shift corrections of as little as 0.1\% can give
substantially different or uncertain mass assignments (Buchler et al. 1996).

 \begin{figure}
 \centerline{\psfig{figure=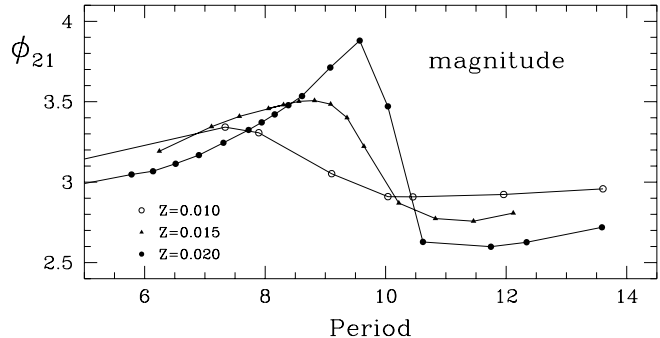,width=10cm}}
 \caption{\baselineskip 0.1cm Behavior of the Fourier phase $\phi_{21}$
as a function of pulsation period for various values of metallicity.
 }
 \label{fig-1}
\end{figure}


\vspace{-2pt}
\subsection{Overtone Cepheids \th (s Cepheids)}
\vspace{-2pt}

The Fourier decomposition parameter $\phi_{21}$ for the first overtone Cepheids
display a "Z" shape around a period of 3 days for the Galaxy (e.g. Antonello et
al. 1990) and around $\approx$ 3.5 and 4.0 days for the LMC (Beaulieu et
al. 1995, Welch et al. 1997) and SMC (Beaulieu \& Sasselov 1997), respectively.
Hydrodynamical models of overtone Cepheids do not agree with the observations
(Antonello \& Aikawa 1993).  Similar results were obtained by Schaller \&
Buchler in a fairly extensive survey of s Cepheids (unpublished preprint,
1994).

\vspace{-2pt}
\subsection{RR Lyrae:}
\vspace{-2pt}

Overall, the modelling of RR Lyrae pulsation gives decent agreement (except for
'double--mode' RRd pulsations) (cf GS for references), but when a more detailed
comparison with observation is carried out in a systematic fashion, serious
discrepancies pop up as Kov\'acs \& Kanbur (1997) show.

\vspace{-2pt}
\subsection{Population II Cepheids}
\vspace{-2pt}

\ni {\sl BL Her stars}:
 The modelling of these low period stars (Hodson, Cox \& King 1982, Buchler \&
Buchler 1994; cf. also GS) shows overall agreement for the lightcurve data, but
the $\phi_{21}$ are considerably smaller than the observational data indicate.

\ni {\sl W Vir and RV Tau stars:}
 It is now well known that the W Vir and RV Tau stars belong to the same group
(Wallerstein \& Cox 1984) and that the properties vary gradually from the low
period, low luminosity W Vir stars to the long period, high luminosity RV Tau
stars.  Although the observations are not very extensive they indicate that the
W Vir stars are periodic up to $\approx$ 15 days from whence they start showing
alternations in the cycles, alternations that become increasingly irregular
with 'period'.  The mechanism for this irregular behavior remained a mystery
until relatively recently.

Numerical hydrodynamical modelling of sequences of W Vir models (Buchler \&
Kov\'acs 1987) uncovered very characteristic nonlinear behavior that goes under
the name of low dimensional chaos.  Since the concept of low dimensional chaos
is still very new in Astronomy we stress that this behavior is very different
from a static multi-periodic, and also different from an evolving multiperiodic
system.  (For the reader familiar with chaos we mention that the presence of
period doubling along sequences of models quite clearly shows the presence of a
horseshoe dynamics with almost one-dimensional return maps.  The chaos in
these models is of the stretch--and--fold type, very similar to the one that
occurs in the R\"ossler system of 3 ODEs, e.g. Thompson \& Stewart (1986).

There is however a discordance with observations, that is the onset of period
doublings and chaos occurs already at 7--10 days, rather than at the $\approx$
15 days indicated by the observations.

\vspace{2pt}

The numerical modelling of RV Tau behavior is much harder because the ratio of
growth-rate/pulsation frequency is much greater.  The pulsations are thus much
more violent and result in sudden loss of the whole envelope in our
calculations (see however Fadeyev \& Fokin 1985, Takeuti \& Tanaka 1995).  We
think that a physical dissipation mechanism is missing from our modelling, even
though we solve the radiation hydrodynamics equations.  Most likely turbulent
dissipation plays a role in taming the pulsations of these stellar models.

\vspace{2pt}

The occurrence of chaos in hydrodynamical models is quite robust as we have
already indicated, but could it be an artifact of the theoretical modelling,
even though it was confirmed with a totally different code (Aikawa 1990).
Clearly it needed to be challenged by observation.  A recent nonlinear analysis
that goes under the name of 'global flow reconstruction' rather conclusively
shows that the irregular lightcurve of R~Sct, a star of the RV~Tau type, is the
result of a low-dimensional chaotic dynamics (For details we refer the reader
to Buchler et al. 1995, or to a didactic review Buchler 1997).  More
specifically, the analysis establishes that the dimension is as low as 4.  In
other words, the lightcurve is generated by 4 coupled ordinary differential
equations.  Put differently, if $s(t)$ denotes the magnitude of the star, then
at any time $t_n$ the lightcurve is a function of {\sl only four preceding
times},
 $s(t_n) = F[s(t_{n-1}),s(t_{n-2}),s(t_{n-3}),s(t_{n-4})]$\hfill\break
 This result is quite remarkable since the pulsations of this star are quite
violent (factors of 40 changes in luminosity!) with shocks and ionization
fronts running about.  As a physicist, though, we are not satisfied merely with
this result, but would like to know what more physics we can learn about this
star.  A four dimensional dynamics indicates that probably two vibrational
modes are involved in the dynamics.  This is strongly corroborated by a
linearization of the dynamics about the equilibrium that tells us that two
spiral stability roots are involved, one unstable with frequency $f_0$=0.0068
d$^{-1}$, the other stable with frequency $f_1$=0.014 d$^{-1} \approxgt$
2$f_0$.  The physical picture that emerges is then that the irregular
lightcurve of R~Sct is the result of the nonlinear interaction between an
unstable, lower frequency mode and a linearly stable overtone with
approximately twice the frequency.

A recent nonlinear analysis of the AAVSO lightcurve of AC~Her similarly shows
low dimensional chaos (Koll\'ath et al. 1997).

To summarize, the predictions of the nonlinear hydrodynamics, viz. that the
irregular behavior of these stars is due to low dimensional chaos, are thus
confirmed, but better numerical modelling is necessary to achieve closer
agreement with observations.


 \begin{figure}
 \centerline{\psfig{figure=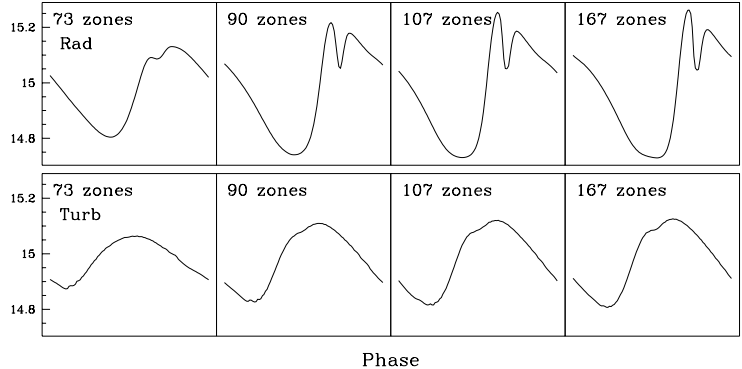,width=10.5cm}}
 \vskip -5pt
 \caption{\baselineskip 0.1cm Behavior of the lightcurve as a function of zone
number with the additional zones between 11,000 and 65,000\th K; top: radiative
adaptive code; bottom: turbulent diffusion code. }
 \end{figure}

Could there be a common cause for most of the "disturbing problems"?  It is
very unlikely that the opacities or the equation of state are still at fault.
A better treatment of radiative transfer (e.g. poster by C. G. Davis) is also
not likely to fix most of the discrepancies.  However, as we have already
pointed out, we seem to be missing a physical dissipation mechanism in our
radiative codes.  In Lagrangean codes it is necessary to include
pseudo-viscosity (\`a la von Neumann-Richtmyer) in order to handle shocks.
While this viscosity is ok for many explosive shock problems, such as supernova
explosions, it unfortunately also provides artificial and unphysical
dissipation elsewhere.  Kov\'acs (1990) found that when he reduced the
artificial dissipation to a minimum the nonlinear pulsation amplitudes kept
increasing to unrealistically large values.  Similarly, when we increase the
spatial resolution of the models the pulsation amplitudes also increase
(Fig.~2).  In fact it turns out that no combination of linear and quadratic
viscosity parameters give satisfactory fundamental and first overtone models.
(We hasten to add though that the Fourier decomposition parameters and the
Floquet stability coefficients of the limit cycles, fortunately, are reasonably
independent of these changes, so that we do not have to throw away everything
we have done so far!)

\vspace{-8pt}
\section{Turbulent diffusion and convection}
\vspace{-5pt}

I had always hoped that, at least near the blue edge of the instability, the
major effect of convection was {\sl static} and would thus merely cause a small
systematic change in the structure of the models, so that we could get away
with purely radiative hydro models.  (Of course it is the important {\sl
dynamic} effect of convection which gives rise to the red edge.)  The problems
and tests that we have described above however indicate that we have to include
turbulent convection in the hydrocodes {\sl in order to provide a missing
powerful dissipation mechanism.}

Turbulent convection is of course a 3D phenomenon and at present, and for some
time to come, it is not possible to run realistic 3D pulsation models.
Progress has been made with relatively idealized 3D convection modelling, but
it is slow and these calculations do not yet allow us to extract the 1D
recipes that we need for our radial pulsation codes.  In the meantime we have
to rely on ad hoc 1D recipes with ad hoc parameters.

The earliest models for convection were local both in time and in space and
were found to be inadequate for stellar pulsations.  Today we have a family of
time-dependent turbulent diffusion models that go back to Spiegel (1963), Unno
(1967) and Castor (1968).  A simplified version was implemented in a
hydrodynamics code by Stellingwerf (1982).  Recent applications have been made
by several groups, viz. Bono et al. (1997), Gehmeyr (1992), Feuchtinger \&
Dorfi (1996) and by Yecko, Koll\'ath \& Buchler (1997).  The strategy has been
and remains to compare the predictions of the models with observations and from
thence calibrate the unknown parameters.

Our own numerical testing shows that with a turbulent diffusion model the
saturation amplitude of the pulsations becomes largely independent of the
zoning (Fig.~2, kindly prepared by Phil Yecko).  Another positive point is that
with the reduced pulsation amplitudes the shocks are absent or much weaker.
This in turn allows one to reduce the artificial dissipation to a very small
value.

As a word of caution, we note though that these models may still be too local
in space (they only have a diffusion operator for the turbulent energy) and the
existence of plumes may have to be taken into account as suggested by Rieutord
\& Zahn (1995).

\vspace{-8pt}
\section{Amplitude Equations}
\vspace{-5pt}

We have already pointed out that the amplitude equation formalism offers an
alternative to 'brute force' numerical modelling.  We would like to stress here
that contrary to the claim of GS this formalism is not an Ansatz, but is a
mathematically rigorous aproach, namely normal form theory.  Essentially, the
only restriction is that the formalism applies to weakly nonadiabatic
pulsators.  Many of the interesting stars, viz. the classical Cepheids, the
RR~Lyrae, the delta Scuti and the white dwarfs definitely fall into that
category.  For details of the formalism as applied to stellar pulsations we
refer the reader to reviews (Buchler 1988, 1993).  We note also that the
formalism has recently been extended to nonradial pulsations, in an Eulerian
formulation by Goupil \& Buchler (1994) and in a Lagrangean one by van Hoolst
(1994).

In a nutshell, the formalism reduces the PDEs of hydrodynamics and radiation
transfer to a small set of ODEs for the amplitudes of the excited modes.  The
structure of the equations is uniquely determined by the types of resonances
that occur among the linear modes of oscillation; the remaining physics is all
contained in the values of the nonlinear coefficients.  The amplitude equations
are generic and capture the essence of the behavior of the system; it is not
astonishing thus that they pop up in many different areas of physics,
chemistry, biology, etc..

The solutions of the amplitude equations (usually the fixed points) tell us
then about the possible types of behavior for a sequence or array of stellar
models.  They also explain the effect of resonances on the morphology of the
lightcurves and radial velocity curves.  Perhaps the most useful and best known
application of the formalism has been to describe the behavior of the Fourier
decomposition parameters through the Hertzsprung progression of the bump
Cepheids (e.g. Buchler 1993).

\vspace{-0pt}
\section{Potpourri}
\vspace{-5pt}

\subsection{'Double-mode' behavior}
\vspace{-2pt}

As an application of the formalism to beat (double-mode) behavior we have
plotted in Fig.~3 the predictions of the amplitude equation formalism in which
it is assumed that two nonresonant modes (e.g. the fundamental and first
overtone) interact and can give rise to a double-mode pulsation along a single
parameter sequence of models (scenario 'AB' in Fig.~1 of Buchler \& Kov\'acs
1986).  Here we denote by $A_0$ both the Fourier amplitude of the fundamental
mode for the fundamental pulsators, and also the amplitude of the fundamental
component in the case of double mode pulsators.  The right figure shows the
corresponding first overtone amplitudes.  {\sl Note that the transition from
single mode to double-mode occurs smoothly} for both amplitudes.  In a
realistic sample of models one would of course get a dispersion both vertically
and horizontally, but the conclusion remains unaffected.

 \begin{figure}
 \centerline{\psfig{figure=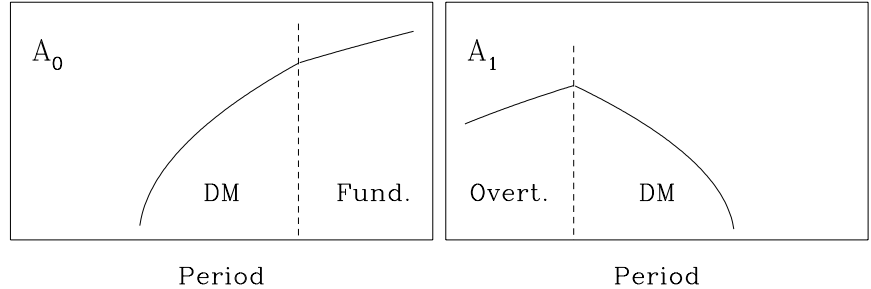,width=12cm}}
 \vskip -1pt
 \caption{\baselineskip 0.1cm Schematic behavior of the fundamental and first
overtone amplitudes for nonresonant scenario; left: fundamental amplitude;
right: first overtone amplitude.
 }
 \end{figure}

 \begin{figure}
 \centerline{\psfig{figure=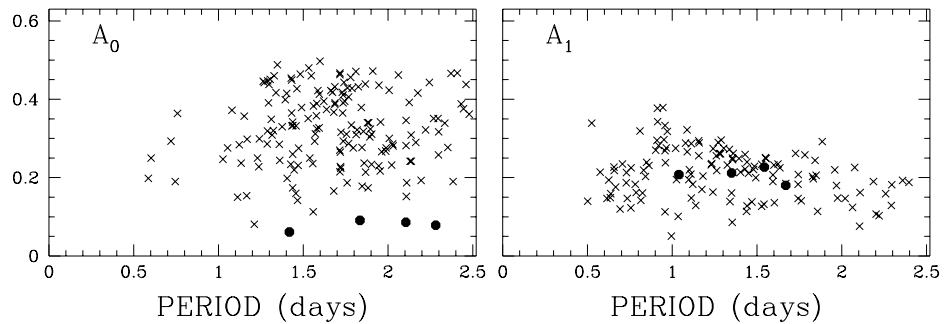,width=13.5cm}}
 \vskip -15pt
 \caption{\baselineskip 0.1cm Behavior of the fundamental and first overtone
amplitudes for the fundamental and first overtone amplitudes of the SMC
Cepheids; left: fundamental amplitude; right: first overtone amplitude
(courtesy of Beaulieu).
 }
 \end{figure}

Let us compare this now to the observations, first to the RR~Lyrae in M15 of
Sandage et al. as reproduced in Buchler \& Kov\'acs 1986.  In their fig.~8 the
first overtone amplitudes are displayed on the left as dots for the RRc and as
crosses for the RRd, whereas the fundamental amplitudes are the dots on the
right for RRab and open circles for the RRd.  The $A_1$ amplitudes of the RRc
and RRd stars indeed vary continuously, but the $A_0$ amplitudes of the RRd are
considerably smaller than those of the RRab.  {\sl We are forced to conclude
that a nonresonant scenario is not in agreement with the observations.  To
explain the jump in the fundamental amplitudes from the RRab to the RRd is is
necessary to invoke the presence of a resonance}.  (A jump in the amplitudes
might also be brought about by higher order, viz. quintic nonlinearities, but
in no studies so far have such nonlinearities ever been found to play a role,
and furthermore the coefficient of the cubic nonlinearity would have to have a
sign opposite to its usual one.)  However, no low order resonances are present
in the stellar parameter range of the RR~Lyrae, and it therefore has to be a
higher order resonance that is involved.  We note in passing that therefore
these stars were better called {\sl beat RR~Lyrae} because more than 2 modes
are involved.

Let us now turn to the Cepheids. J.-P. Beaulieu has kindly provided me with his
SMC Fourier decomposition data that are displayed in Fig.~4.  The first
overtone amplitudes of the beat Cepheids fall right into the range of the
s~Cepheids, but the fundamental amplitudes again are much smaller for the beats
than for the fundamental Cepheids.  We are forced to interpret this to mean
that {\sl a resonance must also be involved in the beat Cepheids}.

\vspace{-2pt}
\subsection{The Blazhko effect}
\vspace{-2pt}

Several mechanisms for the Blazhko effect have been proposed (cf. GS), but a
fully satisfactory understanding has so far defied us.  In the following we
want to present an observational constraint that to our knowledge has not yet
been discussed.  Let us define the Fourier decomposition as $$m(t) = m_o + a\th
cos(\omega t+\phi_1)+b\th cos(2\omega t +\phi_2) + \ldots
 $$
 \ni and as usual $\phi_{21}=\phi_2-2\phi_1$.

 \begin{figure}
 \centerline{\psfig{figure=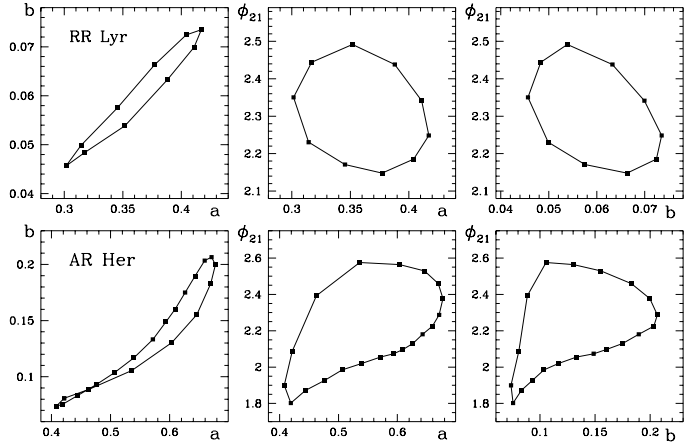,width=10.2cm}}
 \vskip -2pt
 \caption{\baselineskip 0.1cm Variation of the Fourier decomposition parameters
over the Blazhko cycle for RR~Lyrae and AR~Herculis.}
 \end{figure}

The lightcurves over a whole Blazhko cycle have been published by Walraven
(1949) for RR~Lyr and by Bal\'azs \& Detre (1939) for AR~Her.  In Fig.~5 we
show the variation of the pairs of Fourier parameters, a, b, and $\phi_{21}$.
All quantities are seen to oscillate about some center, which is of particular
interest for the phase $\phi_{21}$.  The fact that the phase does not run
through 2$\pi$ over a cycle imposes a severe constraint that can be used to
eliminate some models.

\vspace{-2pt}
\subsection{The dip in the Galactic Cepheid period histogram}
\vspace{-2pt}

In 1977 Becker, Iben \& Tuggle published Cepheid period histograms for several
galaxies.  The histogram for the Galaxy and for M31 showed a pronounced dip in
the 8--10 day period range, whereas the corresponding histograms for the LMC
and SMC were devoid of such a deficiency. Trying to explain the dip on the
basis of their stellar evolution calculations Becker et al. had to invoke an ad
hoc double-humped birthrate function.  Nonlinear calculations show that this is
no longer necessary (Buchler, Goupil \& Piciullo 1997).  Indeed, a perhaps
unexpected side effect of the new opacities is that the fundamental limit cycle
of the Cepheid variables can be unstable.  This instability is found to occur
in the 8--10 day period range for metallicity parameters 0.013 $< Z <$ 0.035.
Note that this is consistent with a dip in the Galaxy and M31 and with the
absence of a dip in LMC and SMC.

\vspace{-2pt}
\subsection{Strange Cepheids and RR~Lyrae}
\vspace{-2pt}

It has recently been found that {\sl strange modes} can occur even in weakly
nonadiabatic stars such as Cepheids and RR~Lyrae.  A thorough study of the
phenomenon has shown that these modes are surface modes that can be
self-excited to the hot side of the blue edge of the normal Cepheid instability
strip (Buchler, Yecko \& Koll\'ath 1997).  The strange modes are recurrent at
higher order wave-vectors, but the lowest ones have typical periods 1/4 to 1/5
that of the fundamental pulsational mode, i.e. they have periods ranging from
$\approx$0.2 days to $\approx$10 days, depending in their luminosity.  Their
locations in schematic HR and PL diagrams are shown in Fig.~6.

 \begin{figure}
 \centerline{\psfig{figure=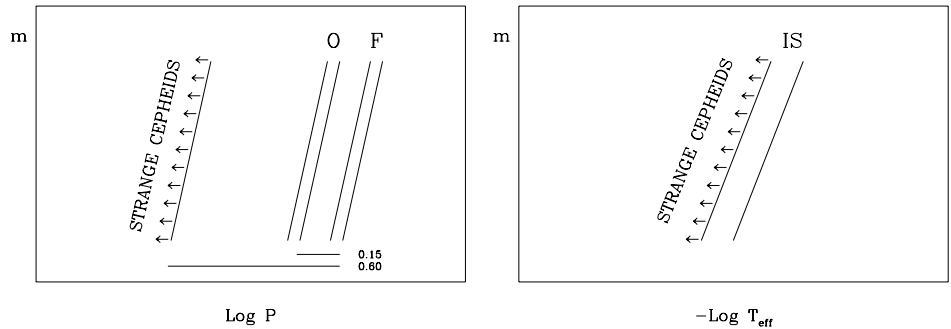,width=13.5cm}}
 \vskip -14pt
 \caption{\baselineskip 0.1cm Schematically, location of strange Cepheids in HR
(left) and P--L (right) diagrams.
 }
\vspace{-5pt}
 \end{figure}

What does one expect the pulsations to look like?  We have computed some
nonlinear (radiative) models and find limit cycles with amplitudes in the
millimag and 10--100 m/s ranges, respectively.  It might be feared that,
because the strange modes are surface modes their driving could be destroyed by
convection.  Preliminary computations with the turbulent diffusion hydro code
however indicate otherwise.

Finally we note that the same trapping and driving mechanisms also work in
RR~Lyrae models and that therefore strange RR~Lyrae should also exist on the
hot side of the RR Lyrae instability range.

\vspace {-8pt}
\section{Conclusions}
\vspace{-5pt}

The theoretical study of stellar pulsations is still faced with many
challenges.  We have seen that radiative hydrocodes, while giving decent
agreement with many observations, are not fully satisfactory.  We hope that a
proper inclusion of the important dissipative effects of turbulent convection
will help resolve many of the extant difficulties and discrepancies.

 \acknowledgments
{\small I wish to congratulate Joyce Guzik and Paul Bradley for the smooth
organization of this excellent meeting, and I also would like to thank them for
their kind financial support.
 This research has been supported by NSF (AST95--18068, INT94--15868) and an
RCI account at the NER Data Center at UF.}

\end{document}